%% file: pozzetti.tex
\documentstyle[11pt,newpasp,twoside,psfig,epsfig]{article}
\def\etal{{et al.\ }}
\def\eblunits{\,{\rm nW\,m^{-2}\,sr^{-1}}}
\def\lta{\mathrel{\hbox{\rlap{\hbox{\lower4pt\hbox{$\sim$}}}\hbox{$<$}}}}
\def\gta{\mathrel{\hbox{\rlap{\hbox{\lower4pt\hbox{$\sim$}}}\hbox{$>$}}}}
\def\sfrd{\,{\rm M_\odot\,yr^{-1}\,Mpc^{-3}}}

\def\edcomment#1{\iffalse\marginpar{\raggedright\sl#1\/}\else\relax\fi}
\reversemarginpar

\begin{document}
\title{The Optical Extragalactic Background Light from Resolved Galaxies}
\author{Lucia Pozzetti}
\affil{Osservatorio Astronomico di Bologna, Via Ranzani 1, Bologna, Italy}

\author{Piero Madau}
\affil{Department of Astronomy and Astrophysics, University of
California, Santa Cruz, USA}

\begin{abstract}

We discuss the ultraviolet to near-IR galaxy counts from the deepest 
imaging surveys, including the northern and southern {\it Hubble Deep Fields}.
The logarithmic slope of the galaxy number-magnitude relation is flatter than 
$0.4$ in all seven $UBVIJHK$ optical bandpasses at faint magnitudes, 
i.e. the light from resolved galaxies has converged from the UV to the near-IR. 
Most of the galaxy contribution to the extragalactic background light
(EBL) comes from relatively bright, low-redshift objects 
(50\% at $V_{\rm AB}\lta 21$ and 90\% at $V_{\rm AB}\lta 25.5$).  
We find a lower limit to the surface brightness 
of the optical EBL of about $15\,\eblunits$, 
comparable to the intensity of the far-IR background from {\it COBE} data.
Diffuse light, lost because of surface brightness selection effects,
may add substantially to the EBL. 
\end{abstract}

\begin{figure}
\centerline{\psfig{figure=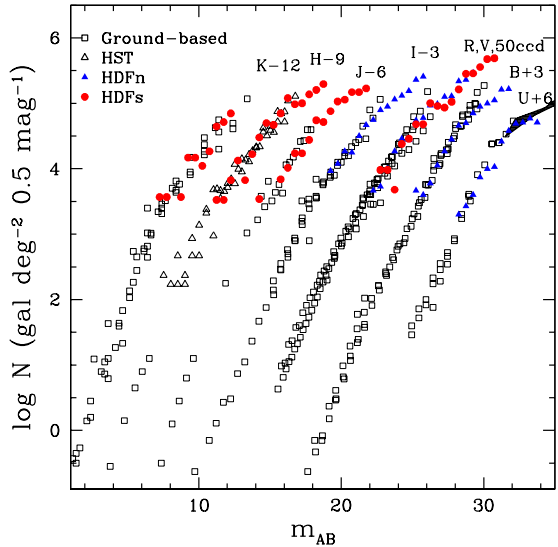,width=8cm}}
\caption{\small Differential $UBVIJHK$ galaxy counts as a
function of AB magnitudes, including {\it HST} and ground-based data.  
Note the decrease of the logarithmic slope $d\log N/dm$ 
at faint magnitudes, with a flattening which is more pronounced at the 
shortest wavelengths.  The shaded region
in the $U$ band shows the results of the ``fluctuation analysis"  by
Pozzetti \etal (1998) (see text).
}
\end{figure}

\section{Introduction}

The extragalactic background light (EBL)
is an indicator of the total luminosity of cosmic structures, 
as the cumulative emission from pregalactic, protogalactic, 
and evolving galactic
systems, together with active galactic nuclei (AGNs), is expected to be 
recorded in this background.
The recent progress in our understanding of faint galaxy data, made possible
by the combination of {\it Hubble Space Telescope} 
{(\it HST}) deep imaging and ground-based spectroscopy, and of the
evolution of the stellar birthrate in optically-selected galaxies from the 
present-epoch up to $z\approx 4$ (Steidel \etal 1999; Madau, Pozzetti,
\& Dickinson 1998),
has been complemented by measurements 
of the far-IR/sub-mm background by the {\it COBE}
satellite (Hauser \etal 1998; Fixsen \etal 1998; Puget \etal 1996), showing  
that a significant fraction of the energy released by stellar nucleosynthesis 
is re-emitted as thermal radiation by dust (Dwek \etal 1998).

In this talk I will focus on the galaxy number-apparent magnitude 
relation and its first moment, the galaxy contribution to the EBL. The 
logarithmic slope of the differential galaxy counts 
($d\log N/dm_{\rm AB}\equiv \gamma$) 
is a remarkably simple 
cosmological probe of the history of the stellar birthrate, as it must 
drop below 0.4 to yield a finite value for the EBL. The radiation emitted from 
unresolved sources that could be lost due to uncertainties in the faintest 
galaxy data and surface brightness selection effects will be discussed,
together with the contribution to the EBL from high-$z$ populations such as
the Lyman-break galaxies and extremely red objects.

\begin{figure}
\centerline{\psfig{figure=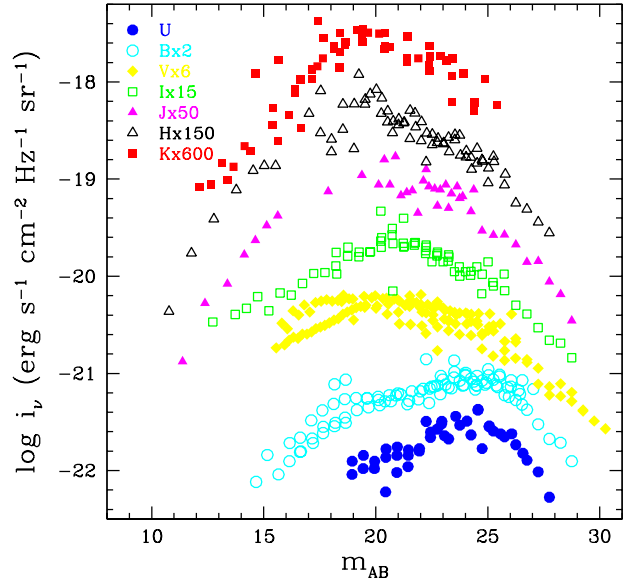,width=8cm}}
\caption{\small Extragalactic background light per magnitude bin,
$i_\nu=10^{-0.4(m_{\rm AB}+48.6)}N(m)$, as a function of $U$ ({\it
filled circles}), $B$ ({\it open circles}), $V$ ({\it filled pentagons}), $I$
({\it open squares}), $J$ ({\it filled triangles}), $H$ ({\it open triangles}),
and $K$ ({\it filled squares}) magnitudes. For clarity, the $BVIJHK$
measurements have been multiplied by a factor of 2, 6, 15, 50, 150, and 600,
respectively.}
\end{figure}

\begin{figure}
\centerline{\psfig{figure=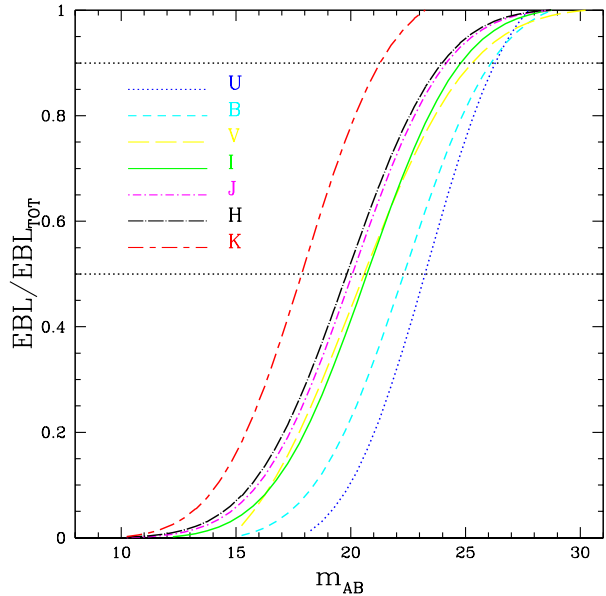,width=7cm}}
\caption{\small Cumulate contribution to the EBL per magnitude bin as
a function of $U,B,V,I,J,H,K$ AB magnitudes, derived from a fit
to the observed counts.}
\end{figure}

\section{Galaxy counts}

Figure 1 shows the northern and southern {\it Hubble Deep Fields} 
(HDFn and HDFs) galaxy counts 
as a function of AB isophotal magnitudes in the $UBVIJHK$ bandpasses.
Details of the data reduction, source detection algorithm, and photometry 
can be found on http://www.stsci.edu/ftp/science/hdfsouth/catalogs.html.
No correction for detection completeness have been made;
in the HDFn the optical counts are likely to be more than $80\%$ 
complete down to the limits plotted in $U,B,V,I$ (Williams \etal 1996). 
A compilation of existing {\it HST} and ground-based data is also shown
(see Madau \& Pozzetti 2000 for references).
All magnitudes have been corrected to the AB system, while the second order
colour corrections for the differences in the filter effective wavelengths 
have not been applied to the ground-based data.

Due to local homogeneity, the differential counts
at bright magnitude %the number-magnitude relation 
follows an euclidean slope ($\gamma \sim 0.6$) 
in all seven band up 
to $m_{\rm AB} \lta 22\div 19$ from $U$ to $K$. 
Because of the curvature of the universe as well as the evolution of galaxies
at intermediate magnitudes, galaxy counts depart from euclidean
expectation and follow a slope  
$\gamma\sim0.45\div0.3$ up to $m_{\rm AB}<26\div 22$ from $U$ to $K$.
At even fainter magnitudes the counts show a clear flattening to a slope 
$\gamma<0.4$ for $m_{\rm AB}>26\div 22$ from $U$ to $K$: in particular
$\gamma<0.2$ in the optical bandpasses $U,B,V,I$,
and $\gamma\sim0.3$ in the $J,H$ and $K$ NIR bands.

A fluctuation analysis by Pozzetti \etal (1998) has shown
that the turnover observed in the $U$ band in the HDFn 
is likely due to the `reddening'
of high redshift galaxies caused by neutral hydrogen along the line of sight.
Recently, a re-analysis of the HDFs by Volonteri \etal (2000) 
shows, however, a steeper slope in the faintest $U$ magnitude bin.
In the B-band the flattening at faint 
apparent magnitudes cannot be due to IGM absorption,
since the fraction 
of Lyman-break galaxies at $B\approx 25$ is small (Steidel \etal 1996; 
Pozzetti \etal 1998).
Moreover, an absorption-induced loss of sources cannot explain the similar
change of slope of the galaxy counts observed in the $V,I,J,H,$ and $K$ bands.

\section{The resolved EBL}

The contribution of known galaxies to the optical EBL can be calculated
directly by integrating the emitted flux times the differential number counts 
down to the detection threshold. 

\subsection{Differential contribution from known galaxies}

The leveling off of the counts is clearly seen in Figure 2, where the function 
$i_\nu=10^{-0.4(m_{\rm AB}+48.6)}N(m)$ is plotted against apparent magnitude in
all bands. 
The differential EBL peaks at $(U,B)_{\rm AB}\sim 24\div25$,
$(V,R,I)_{\rm AB}\sim 21\div22$ and  $(J,H,K)_{\rm AB}\sim 19\div21$. 
While counts having a logarithmic slope $d\log N/dm_{\rm AB}=
\alpha\ge0.40$ continue to add to the EBL at the faintest magnitudes, it
appears that the HDF survey has achieved the sensitivity to capture 
the bulk of the near-ultraviolet, optical, and near-IR extragalactic light 
from discrete sources. 
The effect of counts flattening is depicted in Figure 3, where it is shown 
that relatively bright galaxies contribute the most to the resolved EBL: 
fitting the observed counts we found that 
50\% of it is produced at $(U,B,V,I,J,H,K)_{\rm AB}$ $\lta$ 
(23, 22.5, 21, 20.5, 20, 20, 18), and 90\% at $(U,B,V,I,J,H,K)_{\rm AB}\lta$ 
(26, 26, 25.5, 25, 24, 24, 21.5).

\begin{figure}
\centerline{\psfig{figure=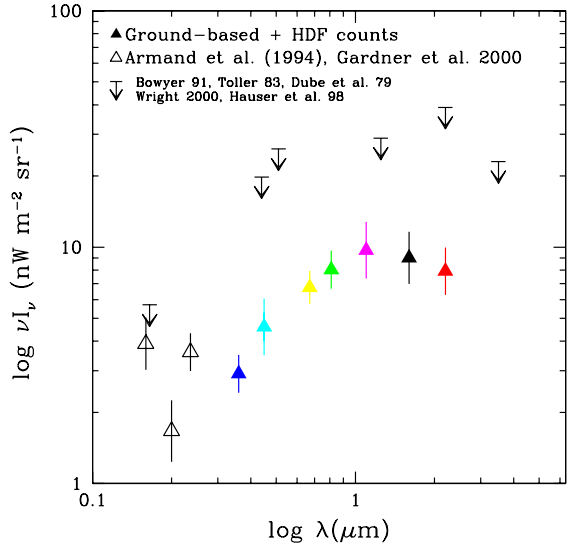,width=7.5cm}}
\caption{\small Spectrum of the optical extragalactic
background light from resolved sources
as derived from a compilation of ground-based and space-based galaxy
counts in the $UBVIJHK$ bands ({\it filled triangles}).}
\end{figure}

\subsection{Spectrum of resolved EBL}

The EBL from detected galaxies in all seven bands has been derived
by integrating the galaxy counts down to the faintest detection threshold.
%we therefore derive 
The results for $0.36\lta \lambda\lta
2.2\,\mu$m are listed in Table 1, along with the 
magnitude range of integration and the estimated 1$\sigma$ error bars, which 
arise mostly from field-to-field variations in the numbers of relatively
bright galaxies.
An extrapolation of the observed $N(m)$
to brighter and fainter fluxes 
would typically increases the integrated light by less than 20\%. 
The integrated galaxy light is 
$\nu I_\nu =2.9 \div 9.7 \eblunits$, increases $\propto \lambda$ from $U$ to 
$I$,
and peaks around $\lambda\sim 1.1 \mu$m.

In Figure 4 we show the spectrum of the integrated galaxy light
including a UV point at
2000 \AA\ from Armand, Milliard, \& Deharveng (1994)
and the results of {\it HST}/STIS integrated counts at 1600 and 2300 \AA\ 
from Gardner, Brown \& Ferguson (2000).
Also plotted are the $90\%$ all-sky-photometry upper limits from Bowyer (1991),
Toller (1983), Dube \etal (1979), Wright (2000), and Hauser \etal (1998).
These are from 3 to 5 times higher than the contribution from known galaxies.

\input table_EBL

\section{The unresolved EBL}
Diffuse, unresolved light may contribute substantially to the optical EBL.
Indeed, different algorithms used for `growing' the
photometry beyond the outer isophotes of galaxies may significantly change
the magnitude of faint galaxies. According to Bernstein (1999) and 
Bernstein \etal (2000), roughly 
50\% of the flux from resolved galaxies with $V>23$ mag lie outside the
standard-sized apertures used by photometric packages. 
An extragalactic
sky pedestal created by the overlapping wings of resolved galaxies also
contributes significantly to the sky level, and is undetectable except
by absolute surface photometry (Bernstein \etal 2000). 
Also, at faint 
magnitude levels, distant objects which are brighter than the nominal depth of 
the catalog may be missed due to the $(1+z)^4$ dimming factor.
All these systematic errors are inherent in faint-galaxy photometry;
as a result, our estimates of the integrated fluxes from resolved galaxies 
will typically be too low, and must be strictly considered 
as {\it lower limits}. 

\subsection{Uncertainties in faint galaxy counts}

An issue in galaxy counts is the uncertain faint slope, particularly in the
$U$ band (cf. Volonteri \etal 2000).
We have estimated the undetected EBL in the case the number counts
have an intrinsic faint slope of $\gamma=0.3$, finding an EBL 
higher by 15$\div$30\% down to zero fluxes. The integrated galaxy contribution 
could increase by 100\%
only if the counts continue to grow with a slope
$\gamma=0.4$ up $m_{\rm AB}\lta35\div40$, or with $\gamma\sim 0.5\div0.8$ up
to $m_{\rm AB}\lta29$, depending on the band.

\begin{figure}
\centerline{\psfig{figure=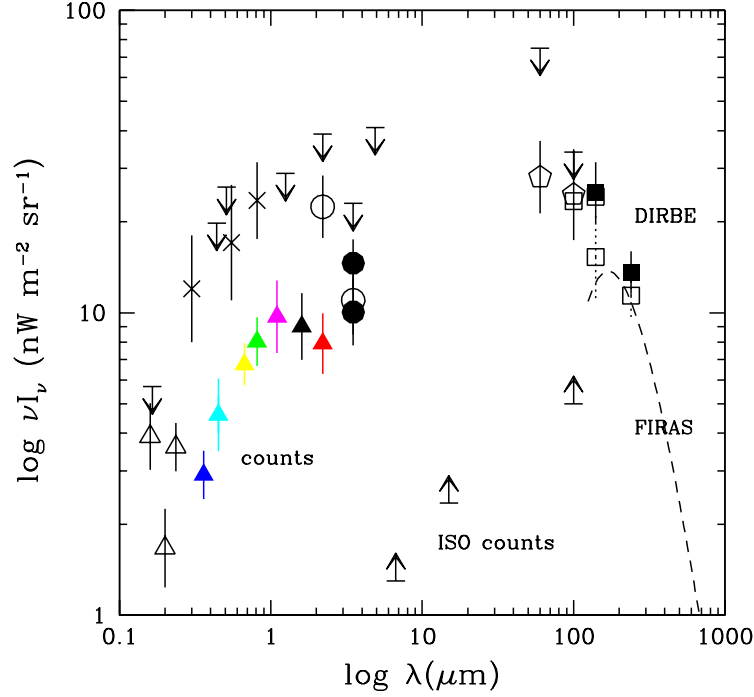,width=10cm}}
\caption{\small Spectrum of the optical extragalactic
background light. The integrated galaxy contribution
as derived from a compilation of ground-based and space-based galaxy
counts in the $UBVIJHK$ bands ({\it filled triangles}) is shown together 
with the FIRAS
125--5000 $\mu$m ({\it dashed line}) and DIRBE 140 and
240 $\mu$m ({\it filled squares}) detections (Hauser \etal 1998; Fixsen
\etal 1998). The {\it empty squares} show the DIRBE points after
correction for WIM dust emission (Lagache \etal 1999, 2000). Also plotted 
({\it empty triangles}) 
is a FOCA-UV point at 2000 \AA\ from Armand \etal (1994) and
the HDF+STIS integrated counts at 1600 and 2300 \AA\ 
from Gardner \etal 2000;
the tentative detection at 3.5 $\mu$m ({\it filled dot}) from
{\it COBE}/DIRBE observations (Dwek \& Arendt 1998), at 2.2, 3.5 $\mu$m from
Gorjian \etal 2000 ({\it empty dot}) and the upper limit from Wright
2000 at 1.25 $\mu$m, and finally
at 60, 100 $\mu$m ({\it empty pentagons}) 
from Finkbeiner \etal 2000.
The crosses at 3000, 5500, and 8000 \AA\ are Bernstein \etal
(2000) tentative measurements of the EBL from resolved and unresolved 
galaxies fainter
than $V=23$ mag (the error bars showing 2$\sigma$ statistical errors).
Upper limits are from Bowyer (1991),
Toller (1983), Dube \etal (1979), and Hauser \etal (1998),
lower limits from ISO counts (Elbaz \etal 1999). }
\end{figure}

\subsection{Surface brightness selection effects}

At faint magnitudes distant objects and low surface brightness 
(LSB) galaxies may be missed due
to the (1+z)$^4$ dimming factor. 
According to Yoshii (1993) 40\% of galaxies are undetected at $B>27$ if
$\mu_{lim}(B)=29$ mag/arcsec$^2$; however, since most of the light comes
from relatively bright objects, the resolved EBL increases only by $\sim20\%$
up to AB$\sim$30. 

The correction for such effects is however model dependent. 
We have estimated it for different scenarios:
from the PLE models of Totani \& Yoshii (2000), 
which include surface brightness selection effects, we found an integrated 
galaxy light which is about 10\% higher in the $B$ band 
up to $B_{\rm AB}<30$, due to high-$z$ objects; 
in a dwarf dominated model (Ferguson \& Babul 1998) the
faint slope of ``total" $I$ counts is 
%the EBL(8100) is
$\gamma\sim0.32$ at $I_{\rm AB}>25.0$; the background from 
galaxies increases then by $\sim15$\% 
due to LSBs and high-$z$ objects.
Vaisanen \& Tollestrup (1999)
estimate that the maximum LSB contribution to the EBL could be 
similar to that from known objects.
Vogeley (1997) shows that there is a uniform unresolved optical
background in the HDFn, 
which would add a fraction from few tens to 50\% to the surface brightness 
from detected galaxies.

Bernstein (1999) has presented
the first tentative detection of the optical EBL at $3000,
5500$ and $8000$ \AA, derived from coordinated data sets from {\it HST} and
Las Campanas Observatory, and shown that the optical EBL 
is a factor 2$\div$4 higher than the integrated contribution of known galaxies.
The EBL in the near-IR has been recently estimated from DIRBE data by 
Gorjian, Wright \& Chary (2000) and by Wright (2000), and shows a similar
excess compared to the integrated galaxy light.
From the above discussion it is difficult, however, to completely 
explain this descrepancy by surface brightness selection effect and flux
lost by standard photometry (Angeretti, Pozzetti, \& Zamorani 2001). 
A high optical/near-IR EBL implies that a high fraction of baryonic mass 
has been processed by stars throughout cosmic history; these stellar baryons
must be accounted for in the local universe (Fukugita
\etal 1998; see also contribution by Madau \etal, this volume).

\section{Optical and FIR EBL}

The spectrum of the optical EBL 
is shown in Figure 5, together with the recent results from {\it COBE}.
The values derived by integrating the galaxy counts down to very
faint magnitude levels 
imply a lower limit to the EBL intensity in the 0.3--2.2 $\mu$m 
interval of $I_{\rm opt}\approx 15\,\eblunits$. Including the tentative
detection at 3.5 $\mu$m by Dwek \& Arendt (1998) would boost 
$I_{\rm opt}$ to $\approx 19\,\eblunits$. 
% if the same correction holds also in the near-IR. 
The {\it COBE}/FIRAS (Fixsen \etal 1998) 
measurements, %yield $I_{\rm FIR}\approx 14\,\eblunits$ 
in the 125--2000 $\mu$m range,
when combined with the DIRBE (Hauser \etal 1998) 
points at 140 and 240 $\mu$m, yield a far-IR 
background intensity of $I_{\rm FIR}(140-2000\,\mu{\rm m})\approx 
20\,\eblunits$.
%The residual emission in the 3.5 to 140 $\mu$m region is poorly known, but it 
%is likely to exceed $10\,\eblunits$ (Dwek \etal 1998). 
The tentative 
direct measurements in the optical 
by Bernstein \etal (2000)  and in the near-IR by Gorjian \etal (2000),
lie between a factor of 2 to 4 higher than the integrated light from 
galaxy counts, with an uncertainty that is largely due to systematic rather than 
statistical error. 
Applying this correction factor to the range 
0.3--3.5 $\mu$m gives a total optical$+$NIR  EBL intensity 
of $\sim 45\,\eblunits$. Including the recent FUV results from Gardner \etal 
(2000),
we derive a `best-guess' estimate of the optical/near-IR EBL intensity 
observed today of 
\begin{equation}
I_{\rm EBL}(0.16<\lambda<3.5 \mu m)=20\div 50\,\eblunits. 
\end{equation}
Gispert \etal (2000) derive a similar value in the FIR,
$I_{\rm EBL}(>6 \mu m)=40\div 52\,\eblunits$.
The FIR/optical EBL ratio ranges between $0.8$ and $2.5$, significanly higher 
than the local value of $0.3$ (Soifer \& Neugeubauer 1991).

\section{Contribution to the resolved EBL at different redshift}

Since the optical galaxy light peaks around $\lambda\sim 1.1 \mu$m 
and the differential resolved EBL shows a maximum at relatively bright 
magnitudes,
most of the background light should come from relatively low
redshift ($z<1$).
In the following we will estimate the contribution to the optical EBL from
two populations of high redshift sources, the Lyman-break galaxies (LBGs) and 
the 
extremely red objects (EROs),
and the predictions of different star formation histories.

\subsection{Lyman-break galaxies}

Using color selection techniques which take into account the opacity at
high redshift of intergalactic matter, 
faint ground-based and {\it HST} observations have made possible the 
detection of star forming galaxies at $z>2$ 
(Steidel \etal 1996, 1999; Madau \etal 1996).
From a statistical analysis of the HDFn Pozzetti et al. (1998) found that
the differential fraction of $z>2$ galaxies ($U$ and $B$ -dropouts) 
increases from $\sim 5$ to $40\%$ in the range $23.5<V_{\rm AB}<27.5$. The 
fraction of U-dropouts to $V_{\rm AB}<27.5$ is $28\pm 2\%$, and of 
B-dropouts to $V_{\rm AB}<28$ is $2.5\pm 0.6\%$,
Integrating the LBG counts and extending them to bright magnitudes using ground 
based observation (Steidel et al. 1999) we estimate a sky brightness 
$I_{z>2}$(V)
$\sim$0.4$\to$1.1 $\eblunits$, which constitutes only a fraction
from 5 to 12\% of the resolved EBL in the visible band.
Near-IR observations have pointed out that LBGs are dusty and therefore must
contribute to the FIR and sub-mm background:
assuming the spectral energy distribution of a star-forming object 
and an amount of dust with E(B-V)$\sim$0.1,  we estimate $I_{\rm FIR}\sim 5
\eblunits$. Adelberger \& Steidel (2000) have argues that LBGs may produce
the entire 850$\mu$m background.

\subsection{Extremely red objects}

The extremely red galaxies (EROs) discovered in deep 
near-infrared (IR) and optical surveys %(e.g., McCarthy et al. 1992, 
(Hu \& Ridgway 1994) 
are defined in terms of their very red optical/near-IR colours ($R-K>5$
or $I-K>4$). While very rare
at bright K magnitudes, their sky density approaches 0.5 $\pm$ 0.1
arcmin$^{-2}$ at K=20 (McCracken et al. 2000).
Such very red colours can be explained by three different scenarios: EROs may
be 1) starburst galaxies hidden by large amounts of dust, or
2) high redshift ($z>1$) old ellipticals with intrinsically red
spectral energy distributions (SEDs) and large positive k-corrections, or 
3) obscured AGNs.
In the last year increasing evidence has been found -- from {\it HST} profiles 
and morphologies (Moriondo \etal 2000; Pozzetti \etal 2001),  
VLT/ISAAC spectra (Cimatti \etal 1999), and clustering properties 
(Daddi \etal 2000) -- that most (up to 70\%) of the EROs could be high-$z$
ellipticals. Some EROs have been detected also in the X-ray by {\it ROSAT} and
{\it XMM} (Hasinger 2000).
Since the space density of EROs is relatively low  up to faint magnitudes 
(Daddi \etal 2000), their
integrated contribution to the EBL in the $K$ band is almost negligible.
At K$<$19.2, Daddi \etal found $(13,2)\%$ of objects with $R-K>(5,6)$
respectively and therefore $I_{R-K>5}(K)\simeq 0.14 
\eblunits \simeq 2\%\,I(K)$ (extrapolating to brighter and fainters 
fluxes the light from EROs converges to 0.3 $\eblunits$, which constitutes 
only about 4\% of the observed EBL in the $K$-band). If, however, EROs are 
mainly dusty starburst at high-$z$, they can contribute to the FIR background:
assuming a star-forming stellar spectrum and a reddening
$E(B-V)\sim0.5\div0.8$, we estimate a contribution to the FIR background of
$\sim 3\div4 \eblunits$.
If EROs are mainly obscured AGNs they may contribute to the hard X-ray 
background.

\subsection{Cosmic star formation history}

\begin{figure}
\centering
\leavevmode
\epsfxsize=.45\hsize \epsfbox {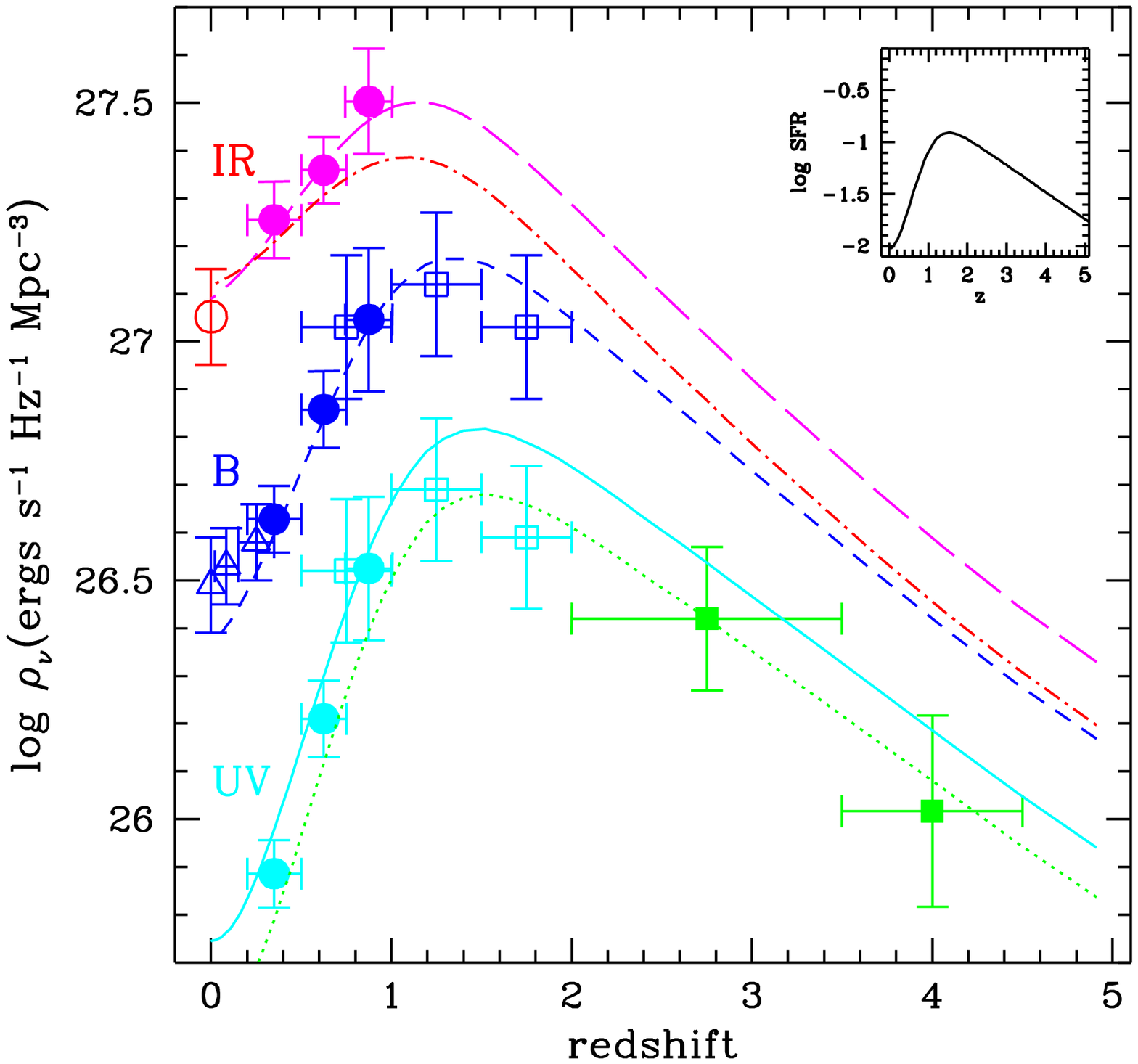} \hfil
\epsfxsize=.45\hsize \epsfbox {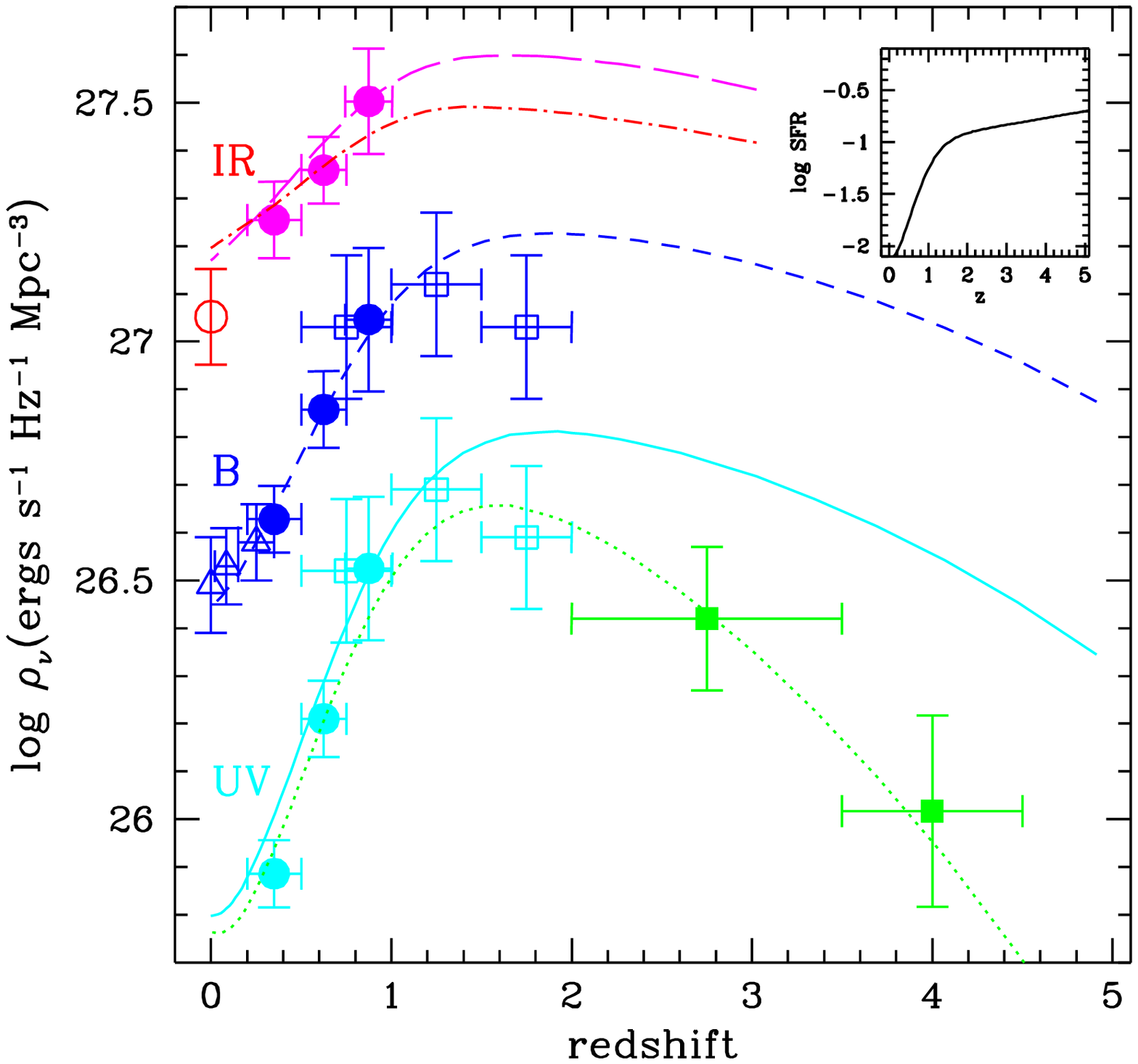} \hfil
\caption{\small
Evolution of the luminosity density at rest-frame wavelengths of 0.15
({\it dotted line}), 0.28 ({\it solid line}), 0.44 ({\it short-dashed line}),
1.0 ({\it long-dashed line}), and 2.2 ({\it dot-dashed line}) \micron\
from Madau, Pozzetti \& Dickinson (1998). The data
points with error bars are taken from Lilly \etal (1996) ({\it filled dots}
at 0.28, 0.44, and 1.0 \micron), Connolly \etal (1997) ({\it empty squares} at
0.28 and 0.44 \micron), Madau \etal (1996, 1998) ({\it filled
squares} at 0.15 \micron), Ellis \etal (1996) ({\it empty triangles} at 0.44
\micron), and Gardner \etal (1997) ({\it empty dot} at 2.2 \micron). The inset
in the upper-right corner of the plot shows the SFR density ($\sfrd$) versus
redshift.  {\it Left panel:} model (A).
{\it Right panel:} model (B) (see text for details).
}
\end{figure}

\begin{figure}
\centering
\leavevmode
\epsfxsize=.4\hsize \epsfbox {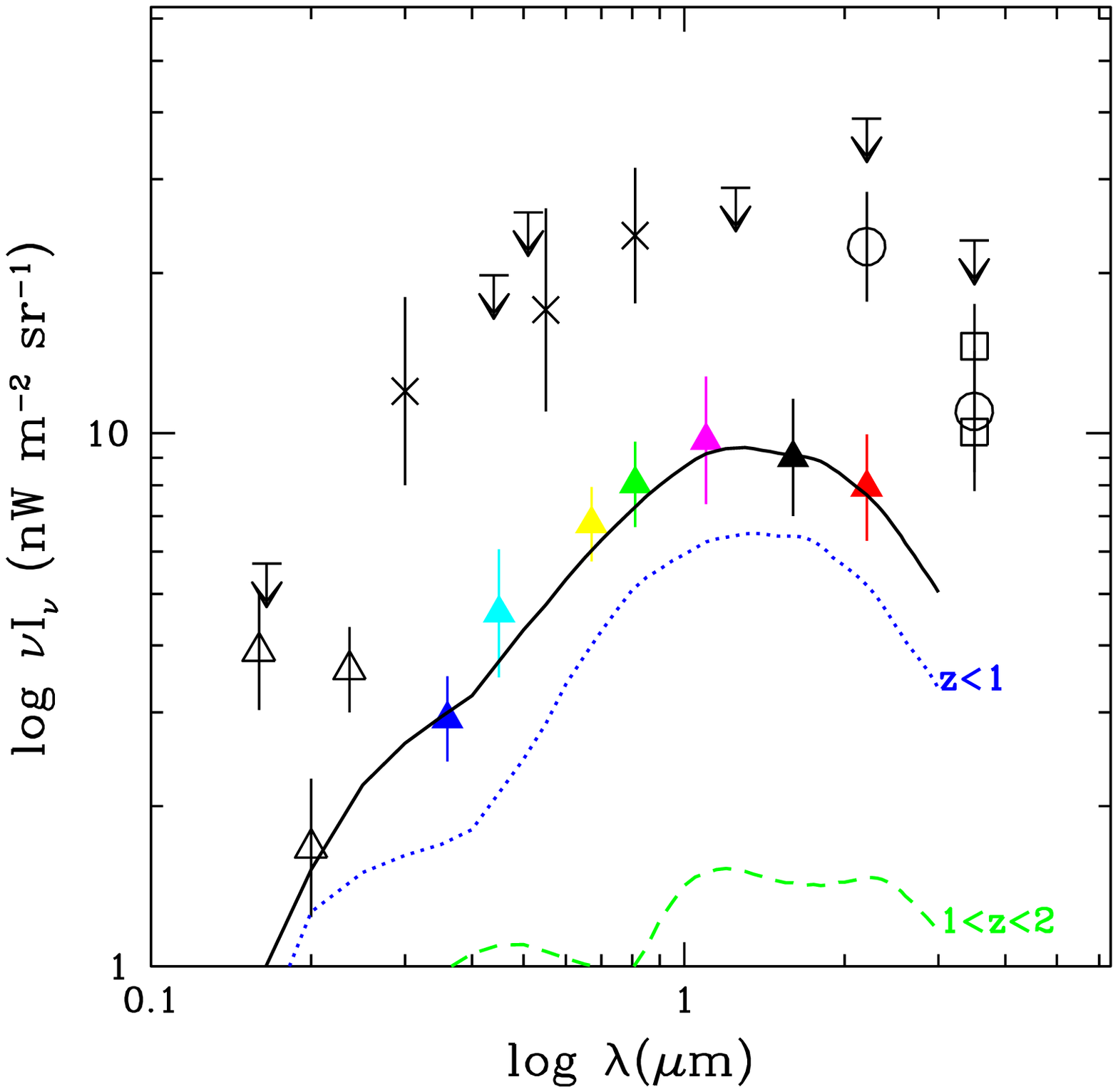} \hfil
\epsfxsize=.4\hsize \epsfbox {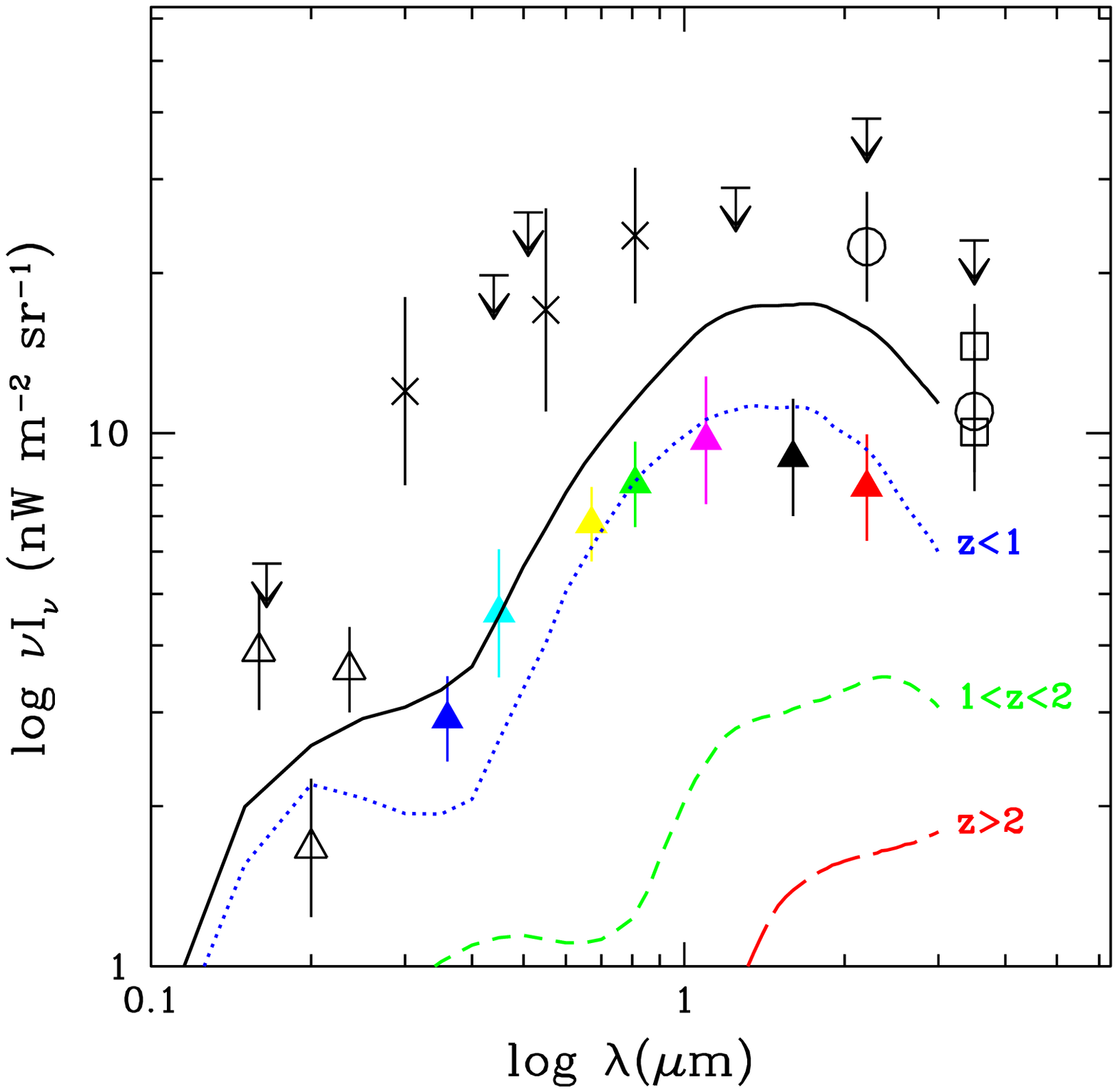} \hfil
\caption{\small Optical EBL produced by model (A) ({\it left panel}) and
model (B) ({\it right panel}) as a function of wavelength ({\it
solid lines}) in different redshift range: $z<1$ ({\it dotted
lines}) $1<z<2$ ({\it short-dashed lines}) and $z>2$ ({\it long-dashed
lines}).
}
\end{figure}

An interesting question arises as to whether a simple stellar evolution
model, defined by a time-dependent SFR per unit volume and a constant IMF, may
reproduce the global UV, optical, and near-IR photometric properties of the
universe as recorded in the comoving luminosity density  as a function
of redshift and in the global EBL.
We assume a universal IMF and fit a smooth function to the UV-continuum
emissivity at various redshifts 
(Lilly \etal 1996; Connolly \etal 1997; Madau \etal 1998; Ellis \etal 1996; 
Gardner \etal 1997).
We then use Bruzual-Charlot's synthesis code to predict the cosmic emission 
history at long wavelengths together with the optical EBL. 

Following Madau, Pozzetti \& Dickinson  (1998), we have constructed two 
simple models: model (A), with a star formation 
history which peaks at $z\sim 1.5$ and decreases at higher redshift,
as expected in a ``hierarchical'' scenario of galaxy formation 
(Somerville \etal 1999) where
about 65\% of the present-day stars formed at $z>1$ and only 20\% at
$z>2$; model (B), with increasing star formation at early times,
as expected in a ``monolithic" scenario where
50\% of the present-day stars formed at $z>2.5$ and were shrouded by dust.
A Salpeter IMF between 0.1 and 120 M$_\odot$ has been assumed in both models
(cf. Madau \& Pozzetti 2000).

\begin{figure}
\centerline{\psfig{figure=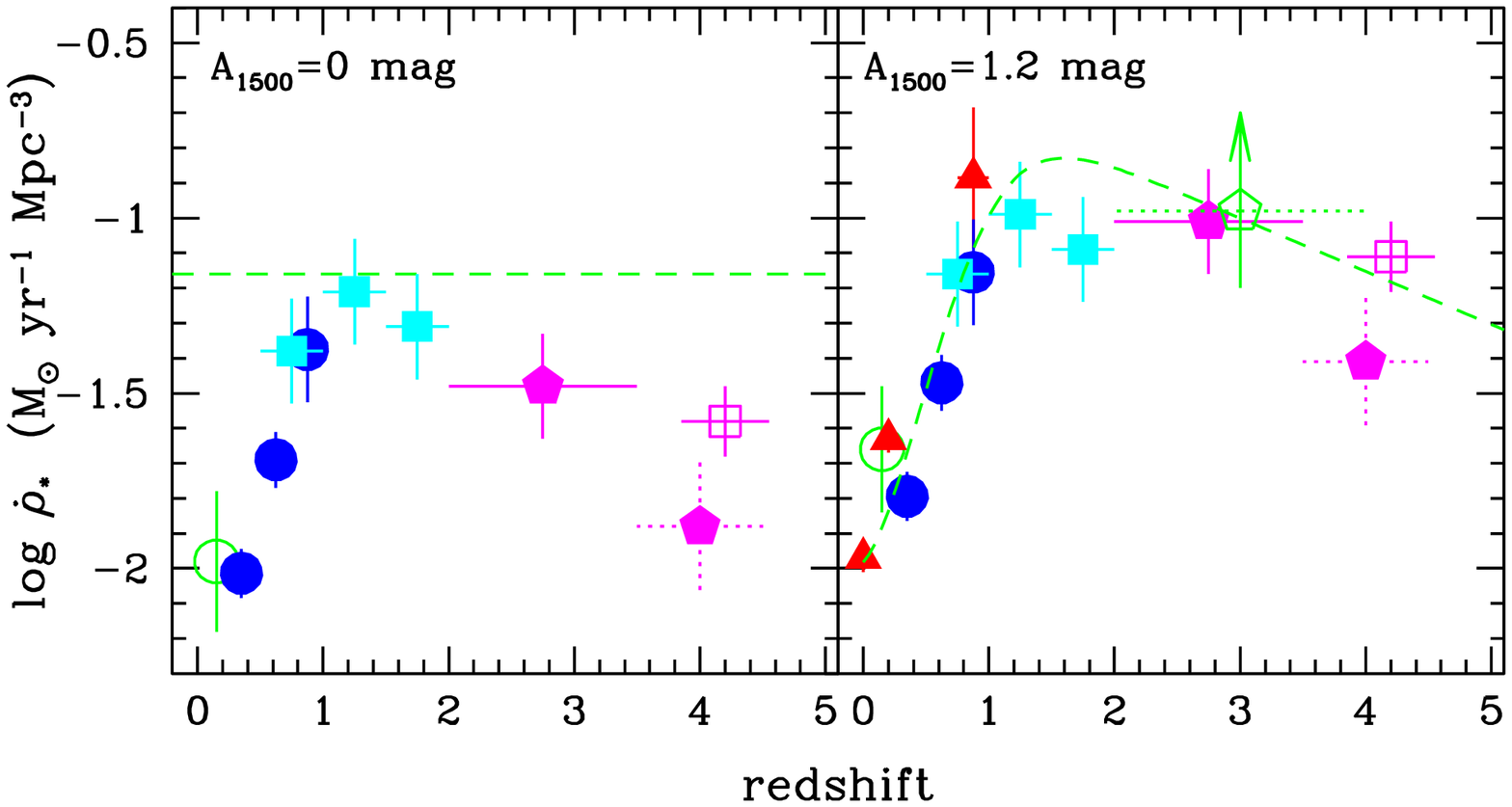,width=12cm}}
\caption{\small  {\it Left:}
Mean comoving density of star formation as a function of cosmic time (from Madau 
1999).
The data points with error bars have been inferred
from the UV-continuum luminosity densities of Figure 6, together with the 
data of Treyer \etal (1998) and Steidel \etal (1999).
The {\it dotted line} shows the fiducial rate, $\langle
\dot{\rho_*}\rangle =0.054\,\sfrd$, required to generate the total EBL. 
{\it Right}: dust corrected values ($A_{1500}=1.2$ mag, SMC-type
dust in a foreground screen). The H$\alpha$ determinations of
Gallego \etal (1995), Glazebrook \etal (1999), and Tresse \& Maddox (1998)
({\it filled triangles}), together with
the SCUBA lower limit (Hughes \etal 1998) ({\it empty pentagon}), have been 
added for comparison.
}
\end{figure}

\begin{figure}
\centering
\leavevmode
\epsfxsize=.45\hsize \epsfbox {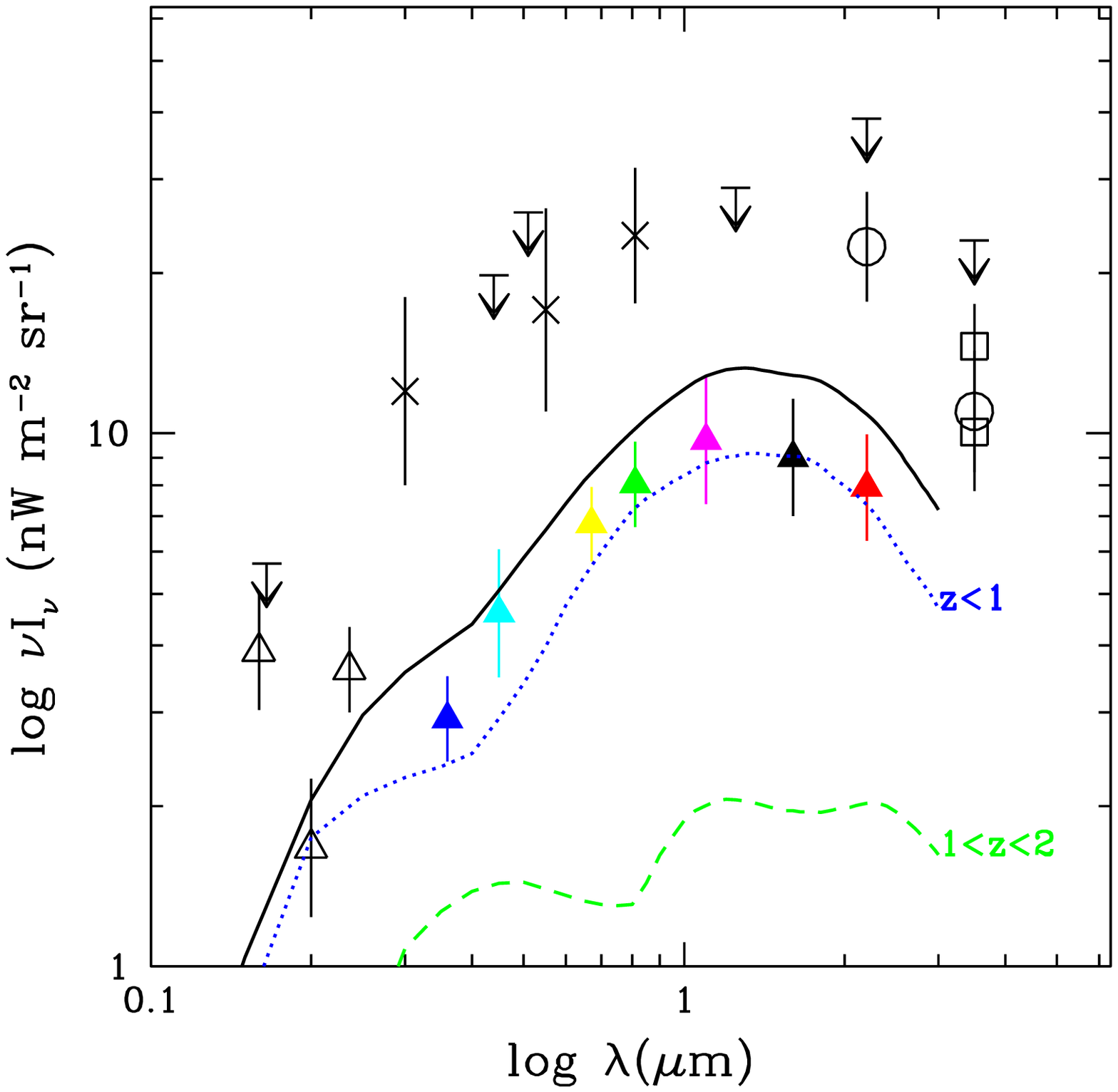} \hfil
\epsfxsize=.45\hsize \epsfbox {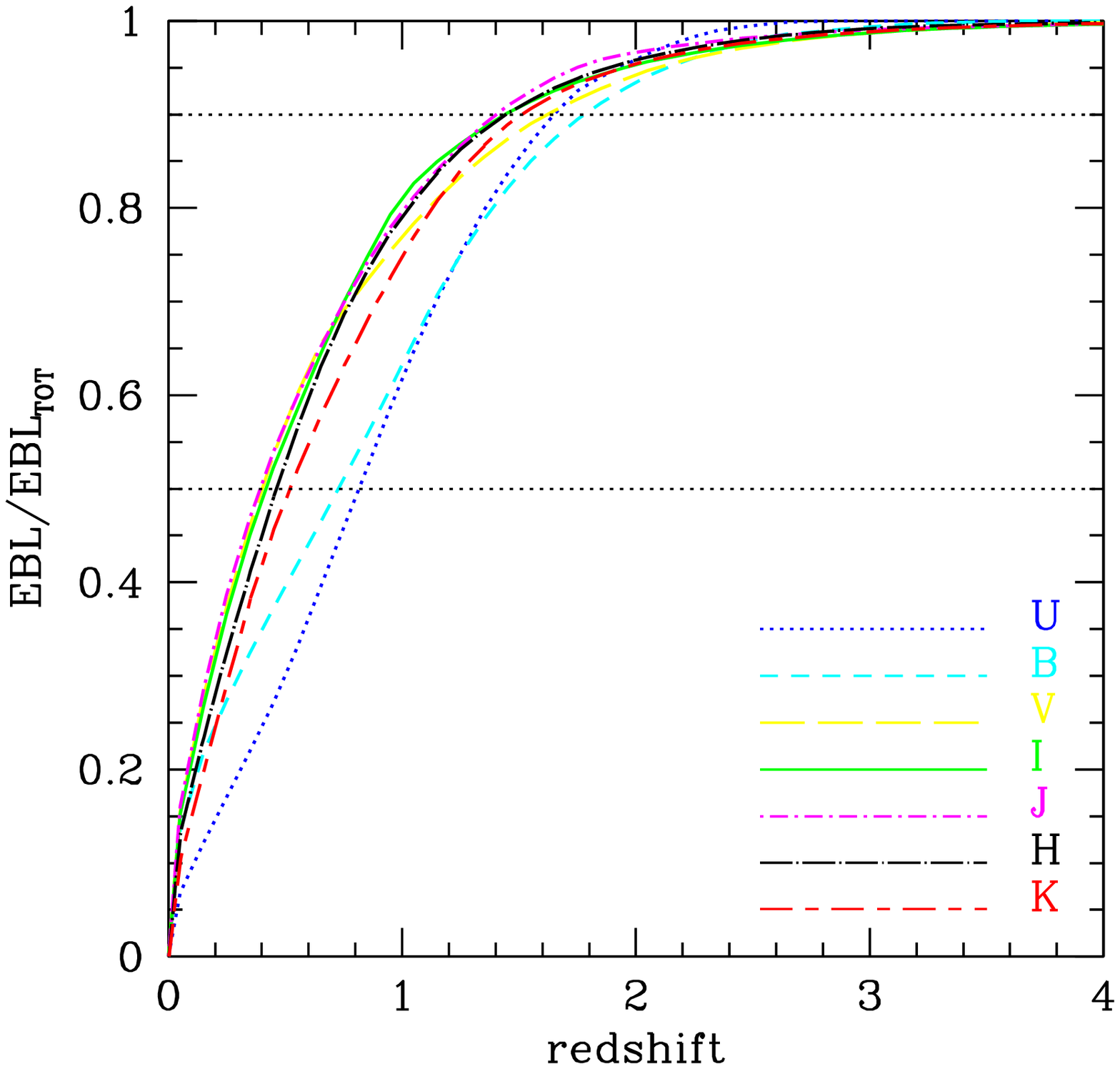} \hfil
\caption{\small 
{\it Right:} Optical EBL for model (C) as in Figure 7.
{\it Left:} Contribution to the optical EBL as a function of redshift
in different bandpasses. 
}
\end{figure}

Figure 6 shows the model predictions for the evolution of the luminosity
density $\rho_\nu$ at
rest-frame ultraviolet to near-infrared frequencies;
the instantaneous star formation rate is shown
in the inset on the upper-right corner of the figure. 
The shape of
the predicted and observed $\rho_\nu(z)$ relations agrees better, within the
uncertainties, in model (A) 
if some amount of dust extinction, $E(B-V)=0.1$, is included.
In this case the observed UV luminosities must be corrected upwards by a
factor of 1.4 at 2800 \AA\, and 2.1 at 1500 \AA. 
In model (B) consistency with the HDF
data has been obtained assuming a dust extinction which increases
rapidly with redshift, $E(B-V)=0.011(1+z)^{2.2}$. 
This results in a correction to the rate
of star formation of a factor $\sim 5$ at $z=3$ and $\sim 15$ at $z=4$.
Overall, the fit to the data is still acceptable, showing how the blue
and near-IR light at $z<1$ are {\it  relatively poor indicators of the star
formation history at early epochs}. 
We have also checked that a larger amount of hidden star formation at 
early epochs, as
advocated by Meurer \etal (1997), would generate too much $B$, 1
\micron\, and 2.2 \micron\ light to be still consistent with the
observations. An IMF which is less rich in massive stars would only exacerbate
the discrepancy. 

The amount of optical starlight produced by the two models
at different wavelengths and redshifts is shown in Figure 7.
While in both scenarios most of the optical light comes from $z<1$ galaxies
at all wavelengths, model (A) recovers the resolved EBL at
all frequencies except the UV. In model (B) $z>1$ galaxies produce additional 
UV and IR light, but there is still a deficit compared with the Gardner \etal 
(2000) resolved UV light and Gorjian \etal (2000) near-IR EBL measurements.
The total amount of starlight radiated at optical wavelengths is
16.5 and 28.6 nW m$^{-2}$ sr$^{-1}$ for model (A) and (B), respectively. By 
comparison, the amount of light absorbed by dust and
reprocessed in the infrared is equal to 9.9 nW m$^{-2}$ sr$^{-1}$ in model
(A), about 38\% of the total radiated flux, while the monolithic collapse
scenario (model B) generates 11.0 nW m$^{-2}$ sr$^{-1}$ 
($\sim 28\%$ of the total flux). 
While both models appear then to be
consistent with the data (given the large uncertainties associated with the 
removal of foreground emission and with the observed and predicted spectral
shape of the CIB), it is clear that much more infrared light would be generated
by scenarios that have significantly larger amount of hidden star formation at 
early and late epochs.

Finally we have used the ``last" version of star formation history corrected
by dust extintion estimated from  
SCUBA observations (Hughes \etal 1998) and near-IR measurements of Balmer lines
(Tresse \& Maddox 1998).
Figure 8 depicts the version of the star formation history with an extintion
correction of $A_{1500}=1.2$ mag (Madau 1999), hereafter model (C), and an 
approximately flat star formation at high-$z$.
In this case the total amount of optical light produced is  23
$\eblunits$, while $\sim 14 \eblunits (\simeq 38\%)$
will be re-radiated in the far-IR.
Also in this case the model reproduces well, within the uncertainties, the EBL 
recorded in the galaxy counts (Figure 9),  while it is still lower than
the detections of Bernstein (1999) and Gorjian \etal (2000).
Half of the resolved EBL in the $(U,B,V,I,J,H,K)$ bandpasses
is radiated at $z<(0.8,0.7,0.4,0.4,0.4,0.5,0.5)$, respectively (Figure 9).

\begin{figure}
\centering
\leavevmode
\epsfxsize=.3\hsize \epsfbox {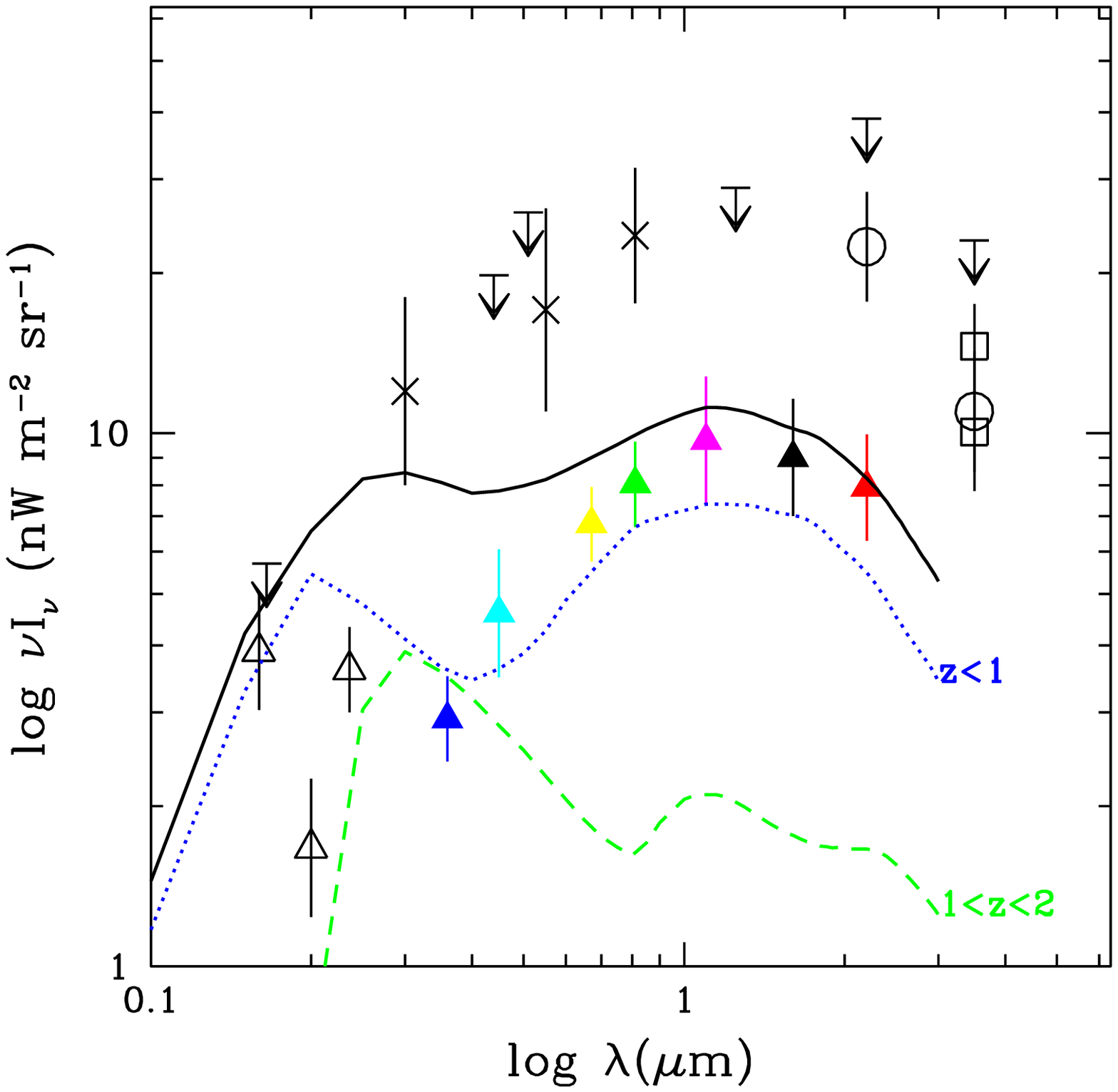} \hfil
\epsfxsize=.3\hsize \epsfbox {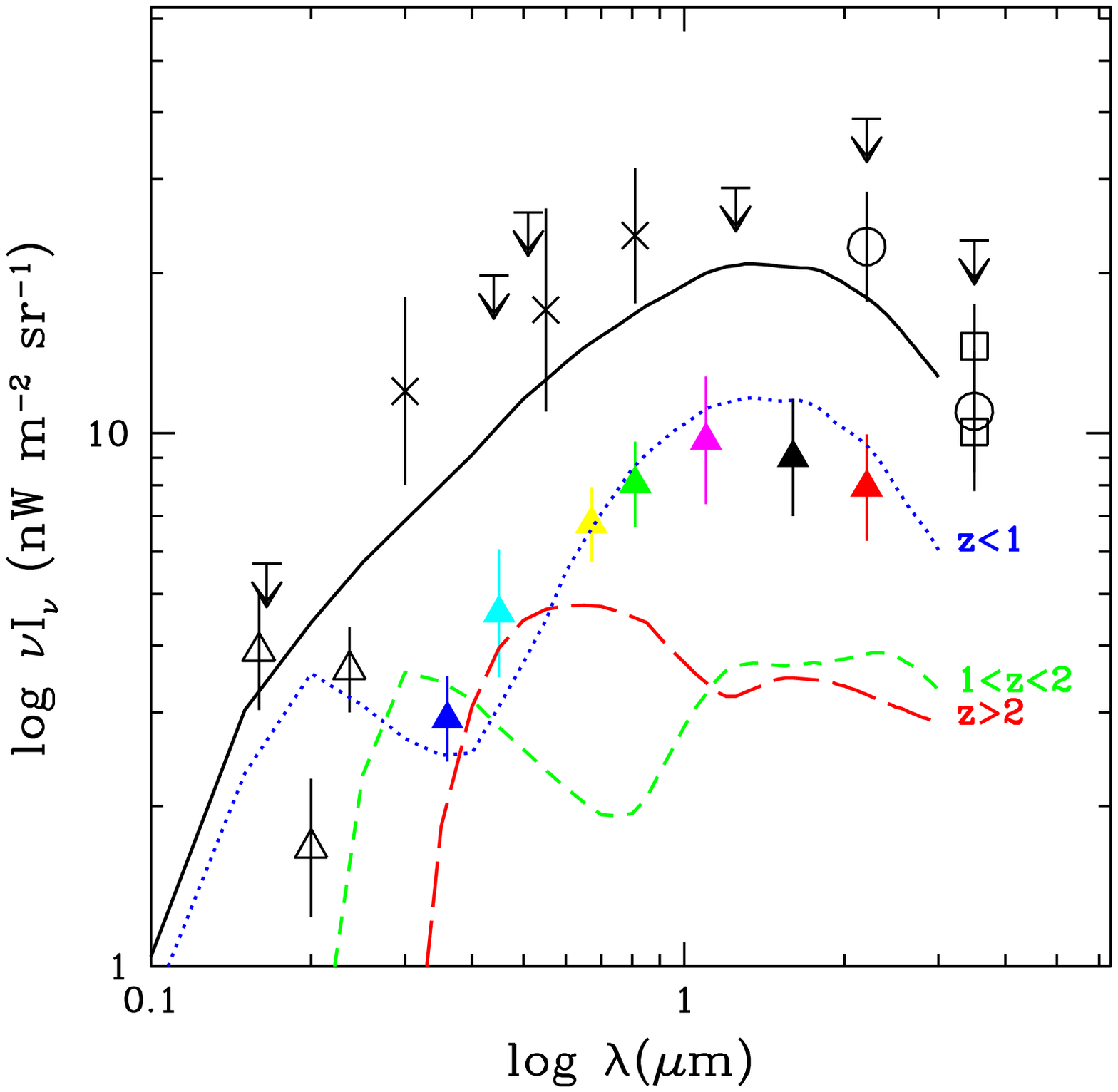} \hfil
\epsfxsize=.3\hsize \epsfbox {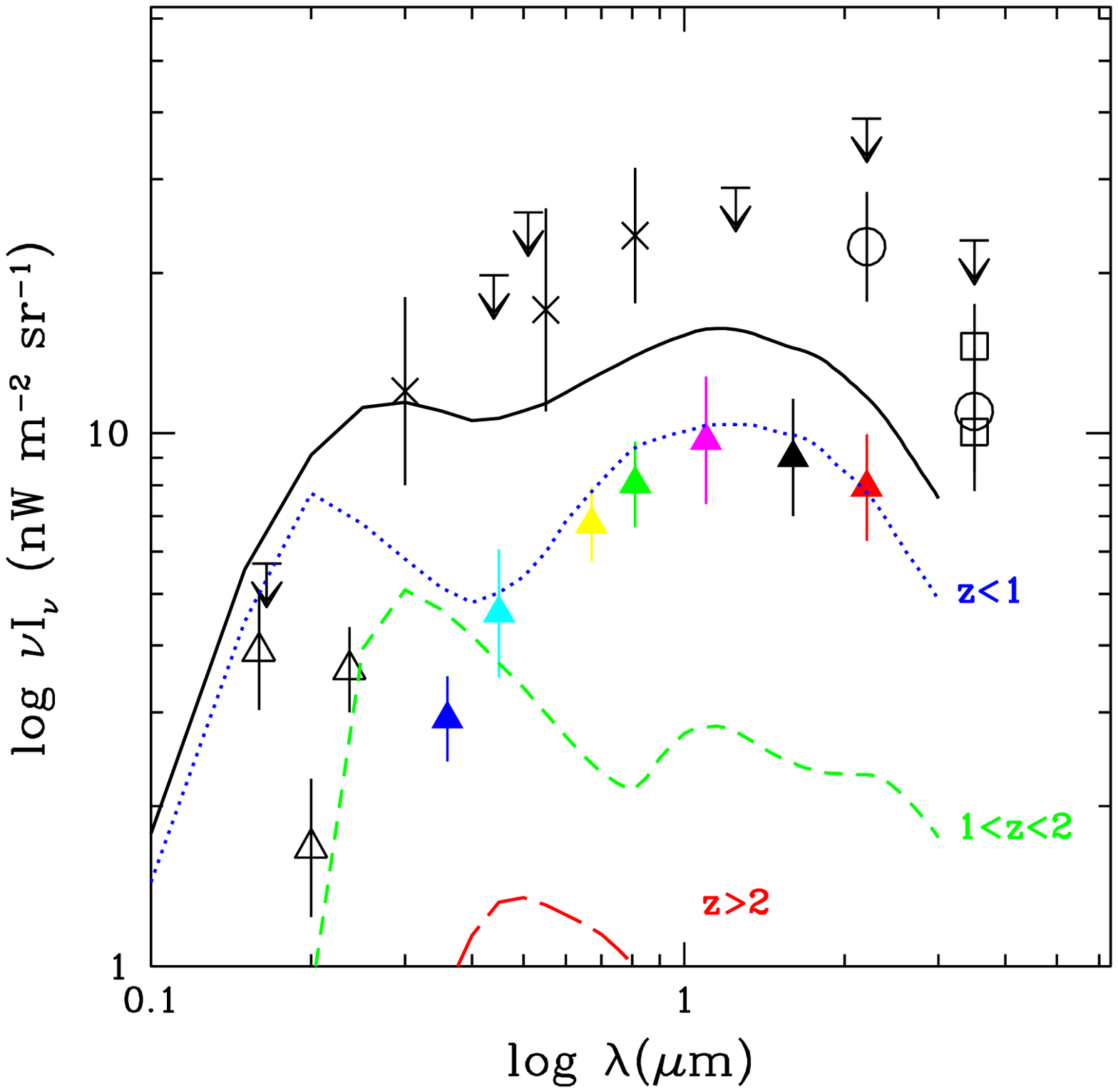} \hfil
\caption{\small 
Optical EBL as in Figures 7 and 9 for the star formation histories of
model (A) ({\it left}), B 
({\it centre}) and C ({\it right}), without dust extinction.
}
\end{figure}

As a test we have compared the EBL predicted by the three models 
without dust extintion.
The results are shown in Figure (10): only in these  unrealistic cases 
we are able to approximately reproduce the EBL detected 
by Bernstein (1999) and Gorjian \etal (2000);
in particular in model (B) and (C) high-$z$ galaxies produce the light
undetected in the galaxy counts at optical and near-IR wavelengths.
In model (B) a non-negligible amount of optical light is radiated at $z>2$.
To account for the far-IR background a higher mean star formation rate must 
assumed, but in this case a larger fraction of baryons must be processed by 
stars 
throughout cosmic history, perhaps in conflict (for the assumed IMF)
with the local stellar census.

\end{document}

%% file: table_EBL.tex
\begin{table}
\centering

\centering{\caption{\small The resolved EBL}}

\vspace{0.2cm}
%{\small Table 1: Integrated Galaxy Light}}
\begin{tabular}{|c|c|c|c|c|} \hline 
%\tablecaption{Integrated Galaxy Light \label{ebl}}
%                &            &               &            &            \\
$\lambda$ (\AA) & AB (range) & $\nu I_{\nu}$ & $\sigma^+$ & $\sigma^-$ \\ \hline
%               &            &               &            &            \\
~$3600$     & $18.0$--$28.0$ & $2.87$ & 0.58 & 0.42 \\
~$4500$     & $15.0$--$29.0$ & $4.57$ & 0.73 & 0.47 \\
~$6700$     & $15.0$--$30.5$ & $6.74$ & 1.25 & 0.94 \\
~$8100$     & $12.0$--$29.0$ & $8.04$ & 1.62 & 0.92 \\
$11000$     & $10.0$--$29.0$ & $9.71$ & 3.00 & 1.90 \\
$16000$     & $10.0$--$29.0$ & $9.02$ & 2.62 & 1.68 \\
$22000$     & $12.0$--$25.5$ & $7.92$ & 2.04 & 1.21 \\ \hline
%\tablecomments{$\nu I_\nu$ is in units of $\eblunits$.}
\end{tabular}

{\small $\nu I_\nu$ is in units of $\eblunits$.}

\end{table}